\documentstyle[prd,aps,preprint,tighten,epsfig]{revtex}

\begin{document}

\draft

\title{Fake $CPT$ Violation in Disappearance Neutrino Oscillations}
\author{\bf Zhi-zhong Xing}
\address{Institute of High Energy Physics, P.O. Box 918 (4), 
Beijing 100039, China \\
({\it Electronic address: xingzz@mail.ihep.ac.cn}) }
\maketitle

\begin{abstract}
We make an analysis of the fake $CPT$-violating asymmetries between
the survival probabilities of neutrinos and antineutrinos,
induced by the terrestrial matter effects, 
in three different scenarios of long-baseline neutrino
oscillation experiments with $L=730$ km, $L=2100$ km and $L=3200$ km.
In particular, the dependence of those asymmetries on
the Dirac-type $CP$-violating phase of the lepton flavor mixing
matrix is examined. 
\end{abstract}


\newpage

It is a real challenge in particle physics to understand the
origin of neutrino masses and the mechanism of lepton flavor 
mixing \cite{Review}. Although the present non-accelerator 
neutrino oscillation experiments have yielded some impressive 
constraints on the lepton flavor mixing parameters \cite{SK,CHOOZ,SNO},
more precise measurements of 
those parameters are desirable and rely on a new generation of 
accelerator neutrino oscillation experiments with long 
baselines \cite{Long}, including the possible neutrino 
factories \cite{Factory}. The terrestrial matter effects 
in such long-baseline neutrino experiments
must be taken into account, since they unavoidably modify the genuine
behaviors of neutrino oscillations in vacuum. A particular manifestation
of the terrestrial matter effects is that they may give rise to  
{\it fake} $CPT$-violating asymmetries between the survival 
probabilities of neutrinos and antineutrinos \cite{Cabibbo}. 

The purpose of this Brief Report is to illustrate the fake 
$CPT$-violating asymmetries between $\nu_\alpha \rightarrow \nu_\alpha$
and $\overline{\nu}_\alpha \rightarrow \overline{\nu}_\alpha$
transitions (for $\alpha = e, \mu, \tau$) in three 
scenarios of long-baseline neutrino oscillation experiments:
$L = 730$ km, $L= 2100$ km and $L= 3200$ km. In particular,
the dependence of those asymmetries on the Dirac-type $CP$-violating
phase of the lepton flavor mixing matrix will be examined. 

Let us consider a transition $\nu_\alpha \rightarrow \nu_\alpha$
(for $\alpha = e$, $\mu$ or $\tau$) and its $CPT$-conjugate process
$\overline{\nu}_\alpha \rightarrow \overline{\nu}_\alpha$ in vacuum.
$CPT$ invariance implies that the probabilities 
of $\nu_\alpha \rightarrow \nu_\alpha$ and 
$\overline{\nu}_\alpha \rightarrow \overline{\nu}_\alpha$ 
transitions amount to each other. Explicitly, we have
\begin{equation}
P(\nu_\alpha \rightarrow \nu_\alpha) \; = \; 1 ~ - ~
4 \sum_{i<j} \left [ |V_{\alpha i}|^2 |V_{\alpha j}|^2 
\sin^2 \left (1.27 ~ \frac{\Delta m^2_{ji}}{E} ~ L \right ) \right ] \; ,
\end{equation}
where $V_{\alpha i}$ (for $\alpha = e, \mu, \tau$ and $i=1,2,3$)
denote the elements of the lepton flavor mixing matrix $V$,
$\Delta m^2_{ji} \equiv m^2_j - m^2_i$ with $m_i$ being the neutrino 
masses (in unit of eV), $L$ is the distance between the production and 
interaction points of $\nu_\alpha$ (in unit of km), and $E$ is the
neutrino beam energy (in unit of GeV). As $T$ conservation is automatic
for the survival probabilities of $\nu_\alpha$ and $\overline{\nu}_\alpha$
neutrinos, $CP$ symmetry holds between 
$P(\nu_\alpha \rightarrow \nu_\alpha)$ and
$P(\overline{\nu}_\alpha \rightarrow \overline{\nu}_\alpha)$ as a 
direct consequence of $CPT$ invariance.

In a realistic long-baseline neutrino experiment,
the genuine pattern of $\nu_\alpha \rightarrow \nu_\alpha$ and 
$\overline{\nu}_\alpha \rightarrow \overline{\nu}_\alpha$ 
transitions is modified due to the presence of terrestrial matter
effects. Defining the {\it effective} neutrino masses $\tilde{m}_i$ 
and the {\it effective} lepton flavor mixing matrix $\tilde{V}$
in matter, one may write out the {\it effective} probability of 
$\nu_\alpha \rightarrow \nu_\alpha$ in the same form as Eq. (1):
\begin{equation}
\tilde{P}(\nu_\alpha \rightarrow \nu_\alpha) \; = \; 1 ~ - ~
4 \sum_{i<j} \left [ |\tilde{V}_{\alpha i}|^2 |\tilde{V}_{\alpha j}|^2 
\sin^2 \left (1.27 ~ \frac{\Delta \tilde{m}^2_{ji}}{E} ~ L \right ) 
\right ] \; ,
\end{equation}
where $\Delta \tilde{m}^2_{ji} \equiv \tilde{m}^2_j - \tilde{m}^2_i$.
The analytically exact expressions of $\tilde{m}^2_i$ and
$\tilde{V}_{\alpha i}$, which depend upon $\Delta m^2_{ji}$, $V_{\alpha i}$
and the matter parameter $A = 2\sqrt{2} ~ G_{\rm F} N_e E$ with $N_e$ being
the background density of electrons and $E$ being the neutrino beam energy,
have been given in Ref. \cite{Xing00}. Note that the {\it effective} 
probability of $\overline{\nu}_\alpha \rightarrow \overline{\nu}_\alpha$ 
can simply be read off from Eq. (2) with the replacements
$V_{\alpha i} \Longrightarrow V^*_{\alpha i}$ and $A \Longrightarrow -A$.
There must be an asymmetry between 
$\tilde{P}(\nu_\alpha \rightarrow \nu_\alpha)$ and
$\tilde{P}(\overline{\nu}_\alpha \rightarrow \overline{\nu}_\alpha)$,
since they are associated respectively with $(V, +A)$ and $(V^*, -A)$.
This {\it fake} $CPT$ asymmetry, denoted as
\begin{equation}
\Delta_\alpha \; \equiv \;
\tilde{P}(\nu_\alpha \rightarrow \nu_\alpha) ~ - ~
\tilde{P}(\overline{\nu}_\alpha \rightarrow \overline{\nu}_\alpha) \; ,
\end{equation}
arises purely from the terrestrial matter effects. A measurement of
$\Delta_\alpha$ is expected to shed light on both the matter
parameter $A$ and the neutrino mixing parameters 
$\Delta m^2_{ji}$ and $V_{\alpha i}$. 

To carry out numerical calculations of the asymmetry 
$\Delta_\alpha$, we adopt the standard parametrization for the
lepton flavor mixing matrix $V$ \cite{FX01}:
\begin{equation}
V \; = \; \left ( \matrix{
c_1 c_3 & s_1 c_3 & s_3 \cr
- c_1 s_2 s_3 - s_1 c_2 e^{-i\delta} &
- s_1 s_2 s_3 + c_1 c_2 e^{-i\delta} &
s_2 c_3 \cr 
- c_1 c_2 s_3 + s_1 s_2 e^{-i\delta} & 
- s_1 c_2 s_3 - c_1 s_2 e^{-i\delta} & 
c_2 c_3 \cr } \right ) 
\left ( \matrix{
1	& 0	& 0 \cr
0	& e^{i\rho}	& 0 \cr
0	& 0	& e^{i\sigma} \cr} \right ) \; 
\end{equation}
with $s_i \equiv \sin\theta_i$ and $c_i \equiv \cos\theta_i$
(for $i = 1, 2, 3$). Since normal neutrino oscillations are completely 
insensitive to the Majorana-type $CP$-violating phases $\rho$ and $\sigma$, 
we only need inputs of the other four parameters 
$\theta_1, \theta_2, \theta_3$ and $\delta$. Indeed 
$(\theta_1, \theta_2, \theta_3)$ are related respectively to the mixing 
angles of solar, atmospheric and reactor disappearance neutrino 
oscillations, at least in the leading-order approximation. 
As the large-angle Mikheyev-Smirnov-Wolfenstein (MSW) 
solution to the solar neutrino problem \cite{MSW} is most favored by the
latest Super-Kamiokande and SNO data \cite{LP01}, we typically 
take $\theta_1 = 35^\circ$ together with 
$\Delta m^2_{21} \approx \Delta m^2_{\rm sun} 
\approx 5\cdot 10^{-5} ~ {\rm eV}^2$. We take $\theta_2 = 40^\circ$ and 
$\Delta m^2_{31} \approx \Delta m^2_{\rm atm}
\approx 3\cdot 10^{-3} ~ {\rm eV}^2$ in view of the present
data on atmospheric neutrino oscillations \cite{SK}, as well as 
$\theta_3 = 6^\circ$ in view of the upper bound 
$\sin^2 2\theta_3 < 0.1$ set by the CHOOZ and Palo Verde 
Collaborations \cite{CHOOZ}.
The size of the Dirac-type $CP$-violating phase $\delta$ remains 
experimentally unrestricted. We consider two
distinct possibilities for illustration: $\delta = 0^\circ$ 
(with vanishing $CP$ violation) and $\delta = 90^\circ$ (with maximal
$CP$ violation). 

If the baseline of neutrino oscillations is about 3000 km or shorter,
it is a good approximation to assume the constant matter density 
of the earth's crust \cite{Barger}: 
$A \approx 2.2 \cdot 10^{-4} ~ {\rm eV}^2 E/[{\rm GeV}]$ with $E$
being the neutrino beam energy. Here let us consider three
interesting scenarios of long-baseline neutrino
experiments: $L = 730$ km, $L = 2100$ km and $L = 3200$ km.
The first baseline corresponds essentially to a neutrino source at the 
Fermilab pointing toward the Soudan mine or that at CERN toward the 
Gran Sasso underground laboratory; the second baseline is compatible
with a possible high-intensity neutrino beam from the High Energy Proton 
Accelerator in Tokaimura to a detector located in Beijing; 
and the third baseline could be a favorable choice for the future 
neutrino factories to measure leptonic $CP$- and $T$-violating 
asymmetries \cite{CP}. 
We are then able to calculate the fake $CPT$-violating asymmetries 
$\Delta_e$, $\Delta_\mu$ and $\Delta_\tau$ with the help of the 
formulas presented already in Ref. \cite{Xing00} and the inputs given above.
Our numerical results are shown in Figs. 1, 2 and 3. 
Two comments are in order.

(a) Different from $\Delta_\mu$ and $\Delta_\tau$, 
$\Delta_e$ is independent of the value of $\delta$.
This feature can easily be understood as follows. 
$\tilde{P}(\nu_e \rightarrow \nu_e)$ relies on 
$\tilde{V}_{e1}$, $\tilde{V}_{e2}$ and $\tilde{V}_{e3}$,
which are in turn dependent on $V_{e1}$, $V_{e2}$ and $V_{e3}$
but independent of the other matrix elements of $V$ \cite{Xing00}.
Since $V_{e1}$, $V_{e2}$ and $V_{e3}$ have nothing to do with $\delta$ in
the afore-mentioned standard parametrization, 
$\tilde{P}(\nu_e \rightarrow \nu_e)$ is completely independent of
$\delta$. A similar argument holds for 
$\tilde{P}(\overline{\nu}_e \rightarrow \overline{\nu}_e)$.
Therefore the asymmetry $\Delta_e$ keeps unchanged for arbitrary
values of $\delta$. 

(b) For the chosen range of $E$ (i.e., 1 GeV $\leq E \leq$ 20 GeV),
the magnitude of $\Delta_e$ at the maximal resonance
can be as large as $2.7\%$ for $L =730$ km, $7.5\%$ for $L=2100$ km, 
or $11\%$ for $L=3200$ km. A measurement of $\Delta_e$ is therefore
possible, at least in principle. The sizes of $\Delta_\mu$ and
$\Delta_\tau$ are too small to be detectable for $L=730$ km; but they
can be as large as a few percent at the resonances, if $L \geq 2000$ km
and $\delta \sim 0^\circ$. Therefore the dependence of $\Delta_\mu$ and 
$\Delta_\tau$ on $\delta$ might be detectable.

We find that the behavior of $\Delta_\alpha$ is quite stable, even
if the matter parameter $A$ fluctuates slightly around the originally
chosen value. The results obtained above are essentially unchanged if
$\Delta m^2_{21}$ flips its sign, but they are sensitive to the 
sign flip of $\Delta m^2_{31}$. In particular, the relation
$\Delta_e (-\Delta m^2_{31}) \approx - \Delta_e (+\Delta m^2_{31})$
is found to hold approximately. Therefore the measurement of 
$\Delta_a$ may help determine
the relative sizes of $m_2$ and $m_3$, which are important to fix the
neutrino mass spectrum. We have also noticed that 
$\Delta_\alpha$ is sensitive to $\theta_3$ \cite{BBB}. 
If $\theta_3$ were too small (e.g., $\theta_3 < 1^\circ$), the
asymmetries $\Delta_e$, $\Delta_\mu$ and $\Delta_\tau$ would be
too small to be practically detected. In a variety of phenomenological
models for lepton masses and flavor mixing \cite{Review,FX96},
$\theta_3 \sim \arctan\sqrt{m_e/m_\mu} \approx 4^\circ$ is expected 
to hold. The measurement of $\Delta_\alpha$
may provide a valuable chance to determine or constrain the smallest
flavor mixing angle $\theta_3$ in the lepton sector.

Finally it is worth mentioning that there might be small {\it genuine}
$CPT$-odd interactions in neutrino oscillations \cite {Glashow}, and 
they might be entangled with the fake $CPT$-odd interactions induced
by the terrestrial matter effects. Whether the former can be disentangled 
from the latter in realistic long-baseline neutrino oscillation 
experiments remains an open question and deserves a systematic study.

\vspace{0.4cm}

The author is grateful to L. Wolfenstein and L.Y. Shan for useful
discussions.

\begin{figure}[t]
\epsfig{file=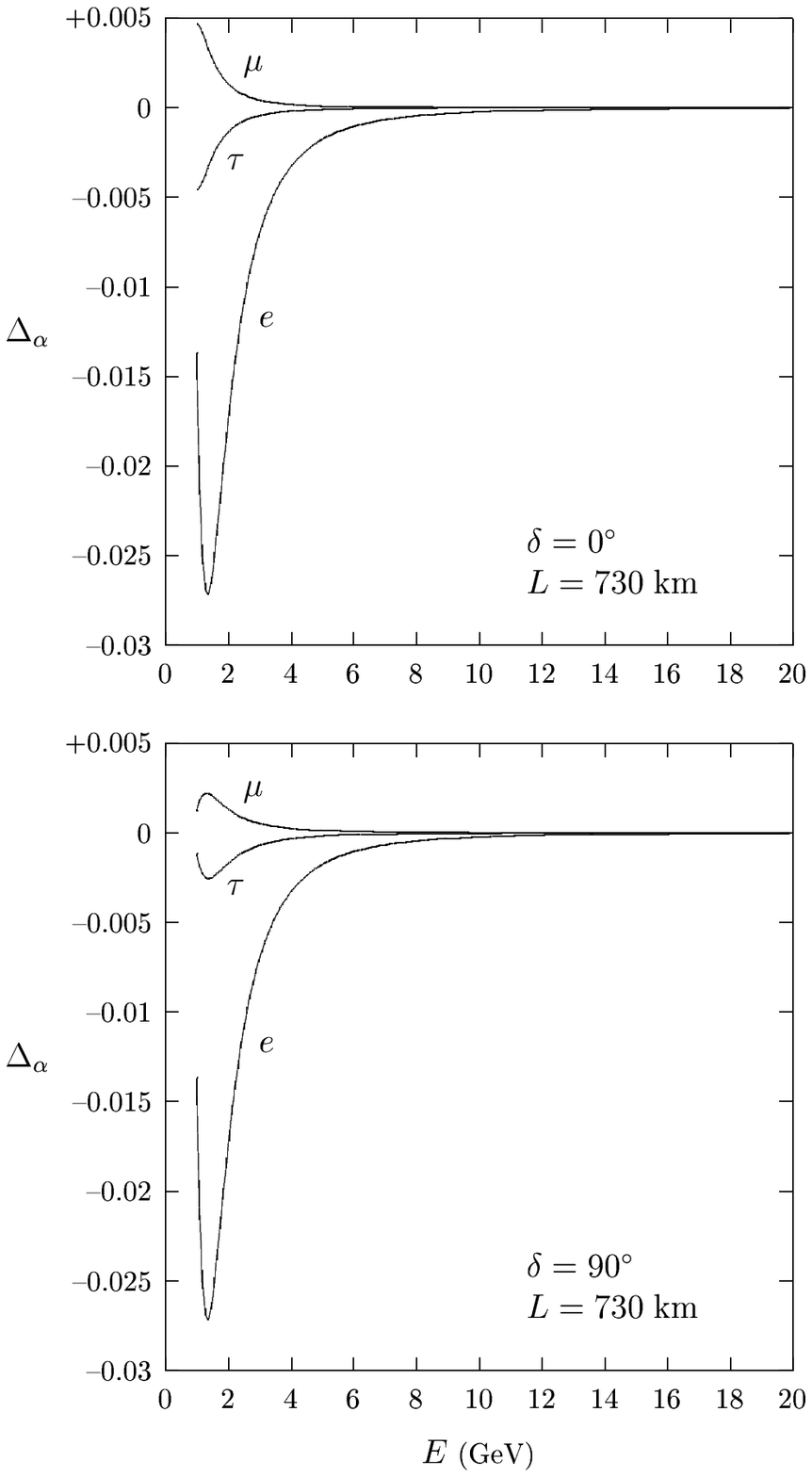,bbllx=-1cm,bblly=1cm,bburx=19cm,bbury=30cm,%
width=15cm,height=20cm,angle=0,clip=}
\vspace{-5.5cm}
\caption{\small Illustrative plots for fake $CPT$-violating asymmetries
$\Delta_e$, $\Delta_\mu$ and $\Delta_\tau$.}
\end{figure}

\begin{figure}[t]
\vspace{-4cm}
\epsfig{file=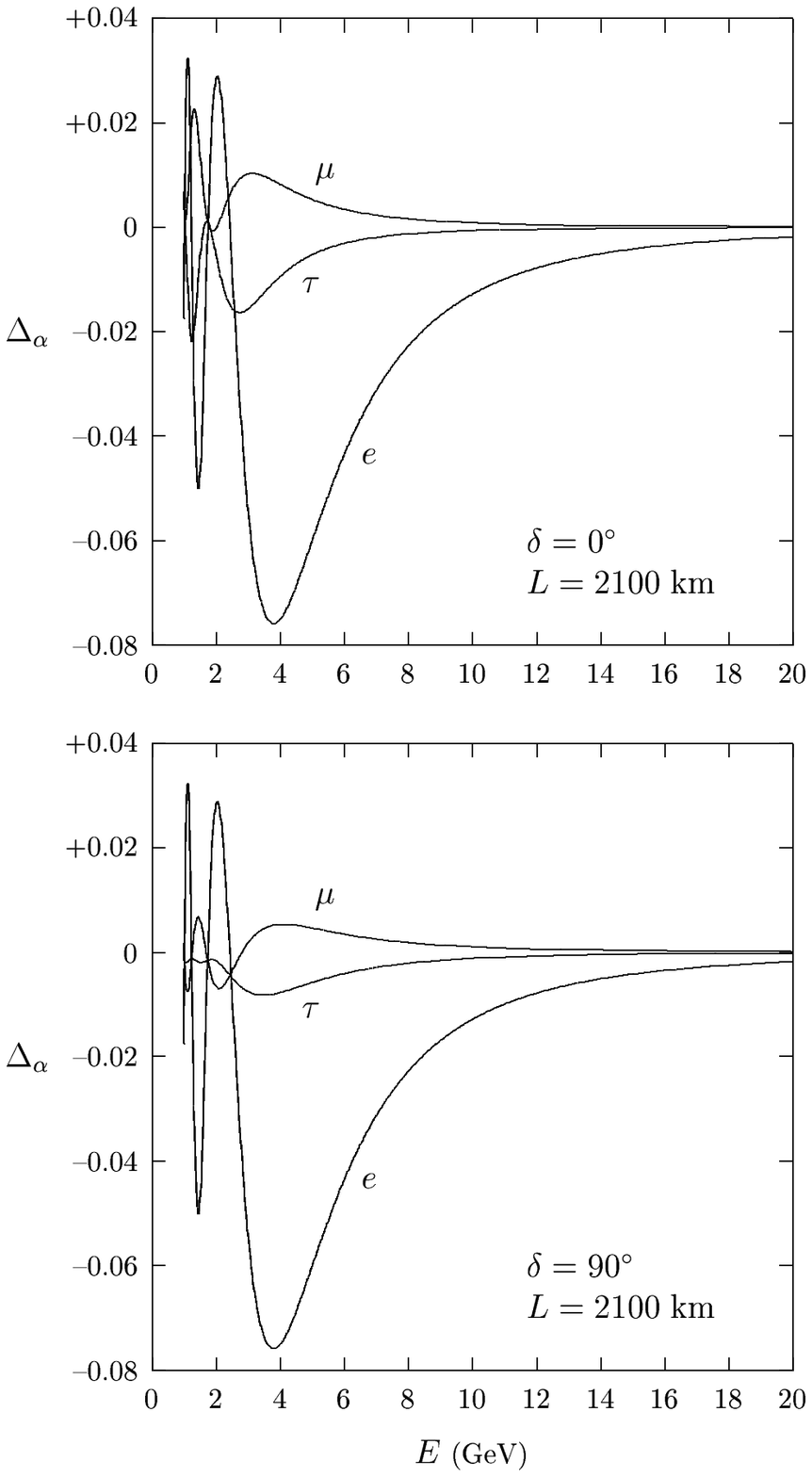,bbllx=-1cm,bblly=1cm,bburx=19cm,bbury=30cm,%
width=15cm,height=20cm,angle=0,clip=}
\vspace{-5.5cm}
\caption{\small Illustrative plots for fake $CPT$-violating asymmetries
$\Delta_e$, $\Delta_\mu$ and $\Delta_\tau$.}
\end{figure}

\begin{figure}[t]
\vspace{-4cm}
\epsfig{file=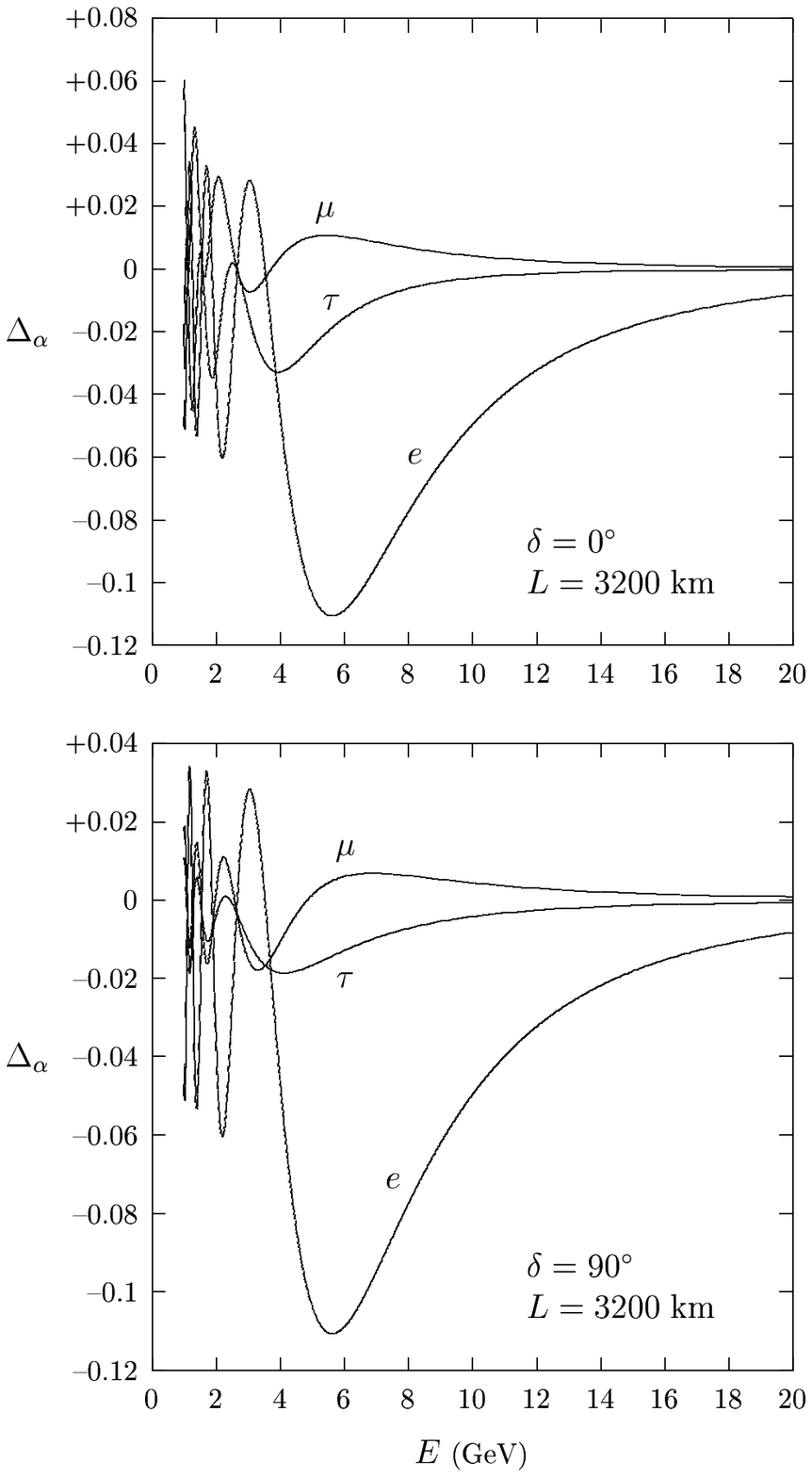,bbllx=-1cm,bblly=1cm,bburx=19cm,bbury=30cm,%
width=15cm,height=20cm,angle=0,clip=}
\vspace{-5.5cm}
\caption{\small Illustrative plots for fake $CPT$-violating asymmetries
$\Delta_e$, $\Delta_\mu$ and $\Delta_\tau$.}
\end{figure}

\end{document}